%% file: main.tex
\documentclass[9pt,twocolumn,twoside]{pnas-new}

\usepackage{lipsum}
\usepackage{bm}                                 
\usepackage{mathtools}                          
\usepackage{subcaption}
\usepackage{siunitx}

\usepackage{xr}
\makeatletter

\newcommand*{\addFileDependency}[1]{
\typeout{(#1)}
\@addtofilelist{#1}
\IfFileExists{#1}{}{\typeout{No file #1.}}
}\makeatother

\newcommand*{\myexternaldocument}[1]{%
\externaldocument[si-]{#1}%
\addFileDependency{#1.tex}%
\addFileDependency{#1.aux}%
}

\myexternaldocument{SI/SI}

\templatetype{pnasresearcharticle} 

\input{definitions}		                        

\title{Nonlinear mechanical metamaterial cloaks}

\author[1]{Giovanni Bordiga}

\author[2,1]{Jean-Gabriel Argaud}

\author[1]{Audrey A. Watkins}

\author[3,1]{Vincent Tournat}

\author[1,*]{Katia Bertoldi}

\affil[1]{John A. Paulson School of Engineering and Applied Sciences, Harvard University, Cambridge, MA 02138, USA}
\affil[2]{MINES Paris -- PSL, Paris, France}
\affil[3]{Laboratoire d'Acoustique de l'Universit\'e du Mans, UMR 6613, Institut d'Acoustique -- Graduate School, CNRS, Le Mans Universit\'e, Le Mans, France}

\leadauthor{Bordiga}

\authorcontributions{
    Author contributions:
    GB, VT, and KB designed research;
    GB and JGA wrote simulation and optimization codes;
    JGA, and AAW performed fabrication;
    JGA, AAW, VT, and GB performed experiments;
    JGA and GB performed simulations and optimizations;
    GB, JGA, AAW, VT, and KB analyzed data and wrote paper.
}
\authordeclaration{The authors declare no competing interest.}
\correspondingauthor{
\textsuperscript{*}To whom correspondence should be addressed. E-mail: \href{mailto:bertoldi@seas.harvard.edu}{\texttt{bertoldi@seas.harvard.edu}.}
}


\begin{abstract}
The concept of cloaking---hiding objects from external detection---has seen wide success in linear systems.
Yet, translating these advancements to nonlinear mechanical systems remains an open challenge.
Here, we present a new approach to nonlinear mechanical cloaking that frames cloaking as an optimization problem aimed at replicating a target mechanical response.
We solve this problem using a differentiable simulation framework coupled with gradient-based optimization.
We implement this approach in a class of mechanical metamaterials constructed from rigid units with elastic couplings that support large deformation and contact interactions.
Using both numerical simulations and physical experiments, we design optimal cloak structures that effectively mask internal inhomogeneities and shield against external mechanical disturbances both in static and dynamic regimes.
This approach provides a versatile design paradigm for creating mechanical systems with integrated cloaking functionality across a broad range of loading scenarios.
\end{abstract}

\dates{This manuscript was compiled on \today}

\begin{document}

\maketitle
\thispagestyle{firststyle}
\ifthenelse{\boolean{shortarticle}}{\ifthenelse{\boolean{singlecolumn}}{\abscontentformatted}{\abscontent}}{}

\section{Introduction}
\label{sec:introduction}
%

Cloaking---the ability to render objects or features undetectable to external probing---has captivated scientific attention across fields.
The concept gained momentum with Pendry's seminal work in optics~\cite{pendry2006controlling,leonhardt2006optical}, where cloaking was achieved through a coordinate transformation technique, now known as transformation optics.
With this transformation, they demonstrated how to route electromagnetic fields around inclusions of arbitrary shape, effectively rendering them invisible to external probing.
This elegant strategy has since been realized experimentally, particularly at the nanoscale in engineered nanostructures~\cite{ergin2010threedimensional,fischer2011threedimensional,ergin2011optical}.
Its influence rapidly extended to other physical domains governed by similar partial differential equations~\cite{martinez2022metamaterials}, including acoustics~\cite{norris2008acoustic,zhang2011broadband,craster2012acoustic}, flexural waves in thin elastic plates~\cite{farhat2009ultrabroadband,stenger2012experiments,colquitt2014transformation,misseroni2016cymatics,golgoon2021transformation}, linear elastostatics~\cite{buckmann2014elastomechanical,buckmann2015mechanical,xu2020physical,sanders2021optimized}, and elastodynamics~\cite{norris2011elastic,quadrelli2021elastic}.
Transformation-based cloaking concepts have also been proposed for diverse applications, such as redirecting seismic waves in seismology~\cite{colombi2016transformation} and ``carpet cloaking,'' which conceals surface irregularities~\cite{ergin2010threedimensional,buckmann2014elastomechanical}.

Recently, new strategies for achieving cloaking have emerged.
In particular, data-driven computational approaches have enabled the design of elastostatic cloaks in lattice metamaterials by optimizing the spatial distribution of unit cells to mitigate the impact of voids or inclusions~\cite{wang2022mechanical}.
Topology optimization strategies have also been successful in designing two-dimensional~\cite{mendez2017computational,cheng2023compatible} and three-dimensional~\cite{senhora2025unbiased} cloaks operating in the linear elastic regime.
In parallel, active cloaking techniques have emerged, employing auxiliary sources to cancel the fields scattered by the object to be concealed~\cite{fleury2015invisible,oneill2015active}.
These active methods have shown considerable promise in acoustic and elastic wave systems, even enabling the experimental realization of ``perfect sensors'' that monitor wave fields without disturbing them~\cite{fleury2015invisible}.

Despite these advances, mechanical cloaking in the nonlinear regime remains unexplored.
The proposed design strategies encounters fundamental roadblocks when extended beyond the linear domain.
Transformation-based methods rely on the invariance of governing equations under coordinate changes---a property that nonlinear elasticity does not possess~\cite{milton2006cloaking,buckmann2015mechanical}.
Active approaches typically depend on Green's functions, which are either unavailable or intractable in nonlinear settings.
Data-driven strategies are typically based on the construction of large databases of precomputed responses.
As such, they are feasible in the linear regime when the response space is low-dimensional, but become computationally prohibitive in the nonlinear regime, where the response depends strongly on the loading path and thorough sampling is no longer practical.

In this work, we propose a design strategy for mechanical cloaking in the nonlinear regime that bypasses the limitations of traditional methods.
By casting cloaking as a \textit{behavior-mimicking} optimization problem---framed as matching the nonlinear mechanical response of a reference system---we eliminate the need for explicit analytical solutions or transformation-based approaches.
We solve this optimization problem using gradient-based optimization within a differentiable simulation framework~\cite{bordiga2024automated}.
Our physical platform consists of mechanical metamaterials made from a network of elastically coupled rigid units capable of large deformations and contact interactions.
These materials have garnered significant attention due to their geometrically tunable nonlinear behavior, enabling functionalities such as programmable stress-strain responses~\cite{deng2022inverse}, multistability~\cite{wu2023situa}, transition waves~\cite{jin2020guided,khajehtourian2020phase}, and complex energy flow manipulation~\cite{bordiga2024automated}.
We systematically design nonlinear cloaking structures that effectively mitigate the influence of both localized excitations and embedded inclusions.
We validate our method through a combination of simulations and physical experiments, demonstrating robust cloaking performance under both static and dynamic loading conditions.
Specifically, we realize metamaterial architectures capable of concealing internal voids, shielding against external forces, and localizing stress within designated regions.
These findings highlight the versatility of our nonlinear cloaking approach across diverse loading scenarios and point toward practical applications in vibration isolation, protection of embedded electronics, and programmable deformations in soft robotics systems.

\section{Design strategy}
\label{sec:design_strategy}
\begin{figure}[htb]
    \centering
    \includegraphics[width=\linewidth]{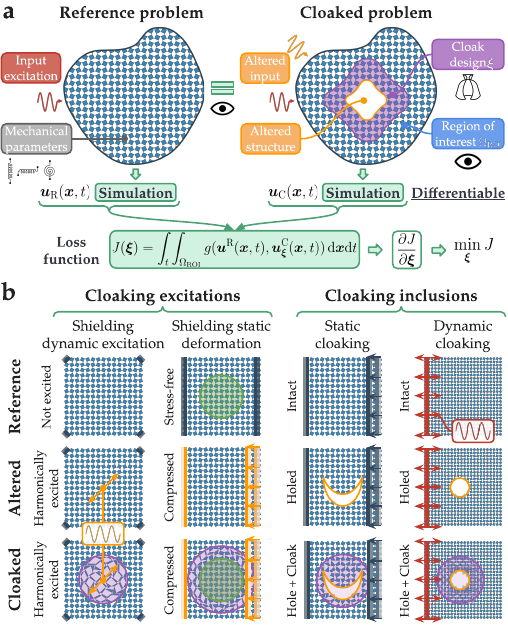}
    \caption{
        \textbf{Design framework for nonlinear mechanical cloaks.}
        (\textbf{a}) Mimicking problem formulation: a reference metamaterial domain (left) and an altered metamaterial domain (right) with a cloak region (purple area) designed to neutralize the effect of the alteration (excitation and/or structural modifications) in a region of interest (blue area).
        Our design strategy is based on the construction of a loss function that quantifies the difference between the reference and altered responses.
        The loss is minimized by tuning the cloak geometry via gradient-based optimization.
        The gradient of the loss with respect to the cloak parameterization $\bxi$ is obtained via a differentiable simulation of the fully nonlinear dynamics of the cloaked problem.
        (\textbf{b}) The proposed framework can address cloaking problems such as providing shielding against external excitations and concealing structural modifications (e.g., inclusions).
    }
    \label{fig:fig_1}
\end{figure}

We consider a metamaterial domain with prescribed geometry, mechanical parameters, and loading (Figure~\ref{fig:fig_1}a, left), and aim to preserve its response within a region of interest despite potential disturbances or modifications.
To this end, we introduce an altered version of the reference problem by changing the structure and/or loading.
Such alterations typically lead to significant deviations in the mechanical response.
Our goal is to locally modify the metamaterial geometry around the altered region to create a cloak that makes the response of the metamaterial in the surrounding region effectively indistinguishable from that of the original reference design (Figure~\ref{fig:fig_1}a, right).

\subsection{Cloaking as a behavior-mimicking optimization problem}
We determine the geometry of the cloak by formulating and solving an optimization problem that enforces response mimicry.
To this end, we assume the ability to accurately simulate the response of both the reference and the cloaked systems.
We denote their respective displacement fields as $\bu^\text{R}(\bx, t)$  and $\bu^\text{C}_{\bxi}(\bx, t)$, respectively, where $\bxi$ is a vector of design variables that defines the geometry of the cloak (see \href{SI-link}{SI}, Section~\ref{si-sec:mathematical_model}).
Here and throughout, $\bx$ and $t$ denote the spatial coordinates and time, respectively.

Given a pair of displacement fields, $\bu^\text{R}(\bx, t)$ and $\bu^\text{C}_{\bxi}(\bx, t)$, we define a non-negative scalar quantity called the \textit{mimicking function}, denoted by $g(\bu^\text{R}(\bx, t), \bu^\text{C}_{\bxi}(\bx, t)) \ge 0$.
In this work, we focus on mimicking either the displacement field or the strain energy of the reference configuration.
For the former, the mimicking function is defined as
\begin{equation}
    \label{eq:g_displacement}
    g(\bu^\text{R}, \bu^\text{C}_{\bxi}) = \norm{ \bu^\text{R} - \bu^\text{C}_{\bxi} }^2 ,
\end{equation}
where $\norm{\cdot}$ denotes the standard $\ell^2$-norm.
Alternatively, to mimic the strain energy, the mimicking function is assumed to be
\begin{equation}
    \label{eq:g_strain_energy}
    g(\bu^\text{R}, \bu^\text{C}_{\bxi}) = \abs{\mW(\bu^\text{R}) - \mW(\bu^\text{C}_{\bxi})} ,
\end{equation}
where $\mW(\cdot)$ denotes the strain energy functional and $\abs{\cdot}$ the absolute value.

We then cast the  cloaking problem as the following optimization problem
\begin{equation}
    \label{eq:general_cloaking_problem}
    \min_{\bxi\in D} J(\bxi) = \min_{\bxi\in D} \int_t \int_{\Omega_\text{ROI}} g(\bu^\text{R}, \bu^\text{C}_{\bxi}) \,\text{d}\bx\text{d}t \,,
\end{equation}
with optimal cloak design $\bxi^*=\arg\min_{\bxi\in D} J(\bxi)$ and $\Omega_\text{ROI}$ being the \textit{region of interest} i.e. the subset of whole domain $\Omega$ where the mimicking effect is desired.
We solve problem defined by Equation~\eqref{eq:general_cloaking_problem} via the Method of Moving Asymptotes (MMA)~\cite{svanberg1987method} with gradient $\text{d}J/\text{d}\bxi$ computed via our differentiable simulation framework~\cite{bordiga2024automated} (see \href{SI-link}{SI}, Section~\ref{si-sec:optimization} for details).
Note that the feasible space $D$ is defined by a set of nonlinear constraints required to ensure manufacturability, such as lower bounds on the edge lengths and angles of the units (see \href{SI-link}{SI}, Section~\ref{si-sec:design_space} for details).

\subsection{Metamaterial platform}
To showcase the proposed approach, we consider flexible 2D mechanical metamaterials composed of networks of rigid units coupled by compliant ligaments.
The nonlinear mechanical behavior of these systems is modeled using a discrete representation, where rigid units are connected at their vertices by elastic springs that can stretch, shear, and bend in a large deformation setting~\cite{bordiga2024automated,coulais2018characteristic}.
A frictionless contact model is also adopted as introduced in~\cite{bordiga2024automated}.
The three degrees of freedom of each unit are denoted by $\bu = \{u_x, u_y, \theta\}$, where $u_x$ and $u_y$ are the horizontal and vertical components of the displacement field, respectively, and $\theta$ is the rotation of the unit.
We derive the Lagrangian of the system and employ automatic differentiation to compute its partial derivatives with respect to all degrees of freedom.
The resulting equations of motion are then solved numerically to simulate the response of the structure (see \href{SI-link}{SI}, Section~\ref{si-sec:equations_of_motion}).

Physical implementation of the metamaterials is realized via 3D-printed polylactic acid (PLA) units connected by thin polyester plastic shims with a rest length of~${\ell_0=\SI{2.3}{mm}}$.
For a detailed description of the fabrication process, see \href{SI-link}{SI}, Section~\ref{si-sec:fabrication}.
The results presented in this work focus on quadrilateral units arranged in a grid-like topology, but the design strategy is agnostic to this choice.
The mass density used in the 2D model is ${\rho=\SI{6.18}{kg/m^2}}$.
The stiffness values of the ligament model are experimentally characterized as ${k_\ell=\SI{120}{N/mm}}$, ${k_s=\SI{1.19}{N/mm}}$, and ${k_\theta=\SI{1.50}{N.mm}}$, with damping coefficients ${c_u=\SI{2.9e-2}{kg/s}}$ and ${c_\theta=\SI{1.2e-7}{kg.m^2/s}}$.
A detailed description of the mechanical parameters and experimental methods is provided in \href{SI-link}{SI}, Section~\ref{si-sec:experimental_methods}.

In all examples presented in this study, the reference design features square units with an initial bias angle of $\theta_0 = \SI{20}{\degree}$ and a centroid-to-centroid spacing of $s = \SI{15}{mm}$ (Figure~\ref{fig:fig_1}b).
For optimization purposes, we also make the degrees of freedom $\bu$ dimensionless with the following normalization
\begin{equation}
    \label{eq:dimensionless_displacement}
    \bU = \{u_x/s, u_y/s, \theta\} \,,
\end{equation}
and refer to $\bU$ as the normalized displacement field.

\section{Results}
By casting cloaking as an optimization problem, we extend the concept to account for large deformations, moving beyond simply neutralizing inclusions to also address challenges such as shielding against external excitations (Figure~\ref{fig:fig_1}b).
In particular, we design nonlinear cloaks that
\begin{enumerate*}[label=(\textit{\roman*})]
    \item block unwanted excitations (Section~\ref{sec:source_shielding}),
    \item generate nearly stress-free regions within a metamaterial domain (Section~\ref{sec:stress_mitigation}), and
    \item conceal void inclusions under both static (Section~\ref{sec:cloaking_inclusion_static}) and dynamic conditions (Section~\ref{sec:cloaking_inclusion_dynamic}).
\end{enumerate*}

\subsection{Shielding dynamic excitations}
\label{sec:source_shielding}
\begin{figure*}[htb]
    \centering
    \includegraphics[width=\textwidth]{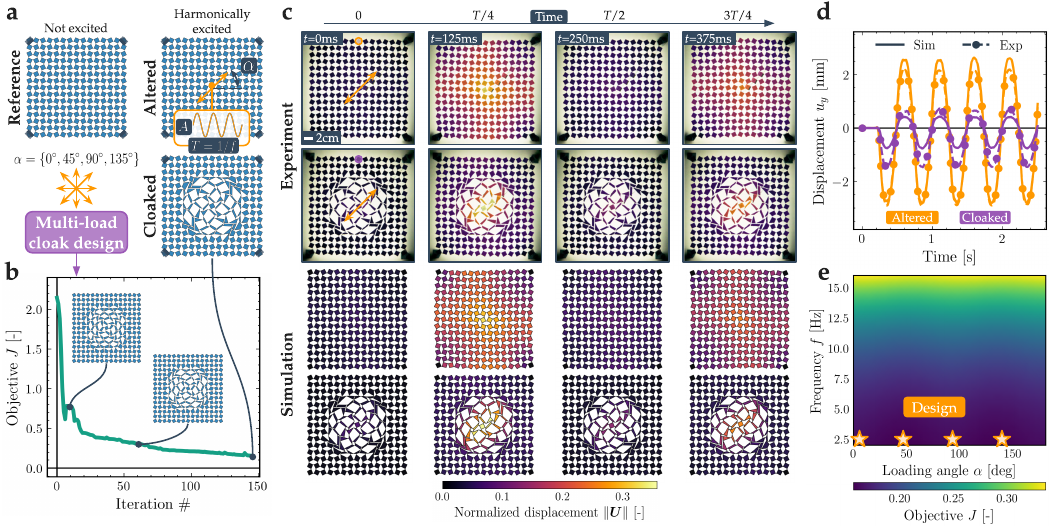}
    \caption{
        \textbf{Shielding dynamic excitations.}
        (\textbf{a}) Cloaking design problem: (top-left) reference structure consisting of an undisturbed $15\times15$-unit rotating-square metamaterial with spacing $s=\SI{15}{mm}$ and angle bias $\theta_0=\SI{20}{\degree}$, (top-right) the altered structure with a time-harmonic excitation applied at the center, and (bottom-right) the cloaked structure with a cloak size of $R_\text{cloak}=4s$.
        (\textbf{b}) Evolution of the loss function during the optimization process for excitation parameters $\alpha\in\{\SI{0}{\degree}, \SI{45}{\degree}, \SI{90}{\degree}, \SI{135}{\degree}\}$, $A=0.3s=\SI{4.5}{mm}$, and $f=\SI{2}{Hz}$.
        (\textbf{c}) Experimental (top) and simulation (bottom) snapshots of the excited reference and cloaked structures over an excitation cycle with period $T=\SI{500}{ms}$ and loading angle $\alpha=\SI{45}{\degree}$.
        Colormap indicates the magnitude of the normalized displacement field $\norm{\bU}$ defined in Equation~\eqref{eq:dimensionless_displacement}.
        (\textbf{d}) Time-evolution of the vertical displacement field $u_y$ in the middle of the top edge of the outer region (orange and purple bullets in \textbf{c}, top-left) over four excitation cycles.
        (\textbf{e}) Cloak performance robustness: contour plot of the loss $J$ as a function of the loading angle $\alpha$ and excitation frequency $f$.
        The orange stars mark the four loading conditions used for the optimization.
        See also Movie~\ref{si-mov:movie_s1} for the dynamic response of both structures under excitations in different directions.
    }
    \label{fig:fig_2}
\end{figure*}
We begin by designing a cloak to suppress unwanted large-amplitude excitations.
The reference configuration consists of an undeformed $15 \times 15$-unit rotating-square metamaterial, with the four corner units clamped to the ground.
This unloaded structure defines the target response we aim to replicate when a time-harmonic excitation, $u_\text{input} = A \sin(2\pi f t)$, is applied to the central unit along a direction defined by the angle $\alpha$ (Figure~\ref{fig:fig_2}a).
In the example considered here, we use an excitation amplitude of  $A=0.3s=\SI{4.5}{mm}$ and a frequency $f=\SI{2}{Hz}$.

Our goal is to design a cloak within a disk of radius $R_\text{cloak} = 4s = \SI{60}{mm}$, centered at the excitation source (Figure~\ref{fig:fig_2}a), that effectively shields the surrounding region from the resulting deformation.
To this end, we solve the optimization problem defined in Equation~\eqref{eq:general_cloaking_problem}, using the displacement-based mimicking function $g$ introduced in Equation~\eqref{eq:g_displacement}.
To ensure robustness with respect to the direction of the applied excitation, we consider four loading angles, $\alpha \in \{ \SI{0}{\degree}, \SI{45}{\degree}, \SI{90}{\degree}, \SI{135}{\degree} \}$, and minimize the average loss across the different loading conditions
\begin{equation}
    \label{eq:source_shielding_loss}
    J(\bxi) = \frac{1}{4} \sum_{i=1}^4 J_{\alpha_i}(\bxi) \,,
\end{equation}
where $J_{\alpha_i}(\bxi)$ is the loss function for the loading angle $\alpha_i$.

Starting from the regular rotating-square pattern, our optimization algorithm rapidly adjusts the geometry of the units within the cloak region to create a structure that effectively shields the outer region from the applied excitation.
After approximately 150 iterations, the algorithm identifies a cloak design that, on average, reduces the magnitude of the normalized displacement in $\Omega_\text{ROI}$ by approximately $70\%$.
The optimized cloak design features a four-fold symmetry, with a low-porosity region on the boundary of the cloak and high-porosity regions in the center as a result of high-aspect ratio units (Figure~\ref{fig:fig_2}a, bottom-right).

To experimentally assess the performance of the optimized design, we fabricate both the reference and cloaked structures and test them under identical loading and boundary conditions, using a low-frequency shaker to apply the excitation (see \href{SI-link}{SI}, Section~\ref{si-sec:setup_testing} for details on the experimental setup).
In Figure~\ref{fig:fig_2}c, we present four experimental and simulation snapshots of both structures when loaded at an angle $\alpha=\SI{45}{\degree}$ (see Movie~\ref{si-mov:movie_s1} for the dynamic response of both structures under excitations in different directions).
The snapshots are color-coded according to the magnitude of the normalized displacement field $\norm{\bU(\bx,t)}$.
The comparison between the excited reference and cloaked structures shows that the cloak functions as a compliant shield, trapping the deformation and significantly reducing the response in the outer region, thereby effectively decoupling the outer region from the excitation source.

To further quantify the shielding effect of the cloak, we monitor the vertical displacement field $u_y$ of the central square unit at the top edge of the samples (marked by the orange and purple circular markers in Figure~\ref{fig:fig_2}c) and present its time evolution over four excitation cycles in Figure~\ref{fig:fig_2}d.
The displacement in the cloaked structure (purple) shows a peak-to-peak displacement of \SI{1.9}{mm}, compared to \SI{5.2}{mm} in the excited reference structure (orange), demonstrating a $63\%$ reduction in the response.
Experiments (dashed-marked lines) and simulations (continuous lines) show a good agreement, with model predictions slightly underestimating the peak-to-peak displacement in the cloaked structure.

Next, we numerically investigate the robustness of the optimized cloak design with respect to the loading angle $\alpha$ and excitation frequencies $f$.
Figure~\ref{fig:fig_2}e presents a contour plot of the loss $J$ as a function of $\alpha \in [\SI{0}{\degree}, \SI{180}{\degree}]$ and $f \in [\SI{2}{Hz}, \SI{16}{Hz}]$, with orange stars marking the four loading conditions used for the optimization.
The results reveal that the optimized cloak effectively shields the outer region across a broad range of angles and frequencies, maintaining a loss below $0.35$ throughout.
This highlights the ability of our design strategy to produce cloaks that remain robust even under conditions far from those considered in the optimization.
\begin{figure}[htb]
    \centering
    \includegraphics[width=0.98\linewidth]{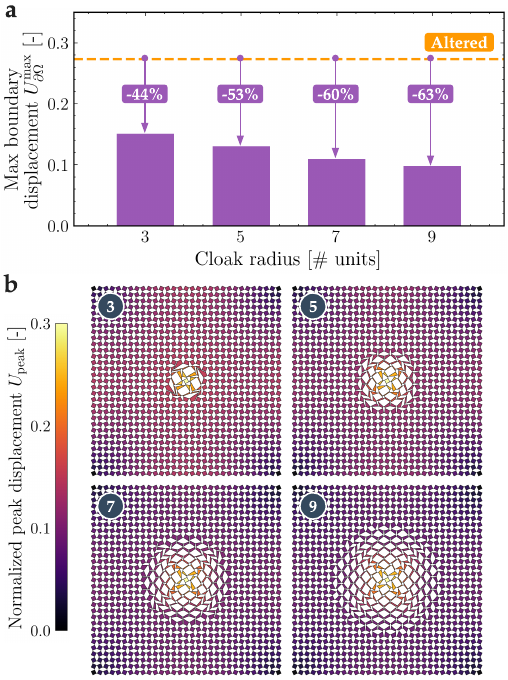}
    \caption{
        \textbf{Effect of cloak size.}
        (\textbf{a}) Cloaking performance as a function of the cloak size $R_\text{cloak}$ for the source shielding problem in $31\times31$-unit rotating-square metamaterials with spacing $s=\SI{15}{mm}$ and angle bias $\theta_0=\SI{20}{\degree}$.
        The performance is quantified by the maximum normalized displacement $U^\text{max}_{\partial\Omega}$ on the boundary of the metamaterial for the cloaked (purple) and altered (orange dashed) structures.
        The cloak is centered at the excitation source, and the region of interest $\Omega_\text{ROI}$ is defined as the area outside the cloak.
        (\textbf{b}) Optimized cloaked structures for different cloak sizes $R_\text{cloak}\in\{3s, 5s, 7s, 9s\}$.
        Colormap indicates the normalized peak displacement field $U_\text{peak}(\bx)$ as given by Equation~\eqref{eq:U_peak}.
    }
    \label{fig:fig_3}
\end{figure}

Finally, we investigate the effect of the size of the cloak, as a larger design space will in general allow for a better optimal cloak geometry.
To this end, we consider a larger metamaterial domain with $31\times31$ units while keeping the same spacing $s=\SI{15}{mm}$ and angle bias $\theta_0=\SI{20}{\degree}$.
We deploy our optimization algorithm to discover cloaks of increasing size $R_\text{cloak}\in\{3s, 5s, 7s, 9s\}$.
To compare the cloaking performance of the different cloak sizes, we compute the maximum normalized displacement on the boundary of the metamaterial as
\begin{equation}
    \label{eq:U_max}
    U^\text{max}_{\partial\Omega} = \max_{t} \max_{\bx\in\partial\Omega} \norm{\bU(\bx, t)} \,,
\end{equation}
and report the results in Figure~\ref{fig:fig_3}a along with the maximum normalized displacement on the boundary of the altered structure (orange dashed line).
Note that the maximum boundary displacement $U^\text{max}_{\partial\Omega}$ is a better metric to compare the performance of different cloak sizes than the loss function $J$, since the latter is influenced by the size of the region of interest $\Omega_\text{ROI}$.
As shown in Figure~\ref{fig:fig_3}a, the maximum displacement monotonically decreases with increasing cloak size while also displaying diminishing returns.
The optimized cloaked structures for $R_\text{cloak}\in\{3s, 5s, 7s, 9s\}$ are shown in Figure~\ref{fig:fig_3}b, with quadrilateral units colored according to the normalized peak displacement field $U_\text{peak}(\bx)$ defined as
\begin{equation}
    \label{eq:U_peak}
    U_\text{peak}(\bx) = \max_{t} \norm{\bU(\bx, t)} \,.
\end{equation}
The spatial distribution of the peak displacement field in Figure~\ref{fig:fig_3}b shows how a larger cloak is more effective at `trapping' the large deformation inside the cloak, thus reducing the overall deformation experienced in the outer region.

It is worth noting that, similar to the smaller cloak design shown in Figure~\ref{fig:fig_3}c, all these optimized configurations exhibit a cross-like structure oriented at $\SI{45}{\degree}$, dividing the system into four radial sectors filled with slender units.
As the cloak size increases, a clear trend emerges: the slender units arrange into a pattern of alternating large and small, approximately rhombic voids.

\subsection{Shielding quasi-static deformation}
\label{sec:stress_mitigation}
\begin{figure*}[htb]
    \centering
    \includegraphics[width=\textwidth]{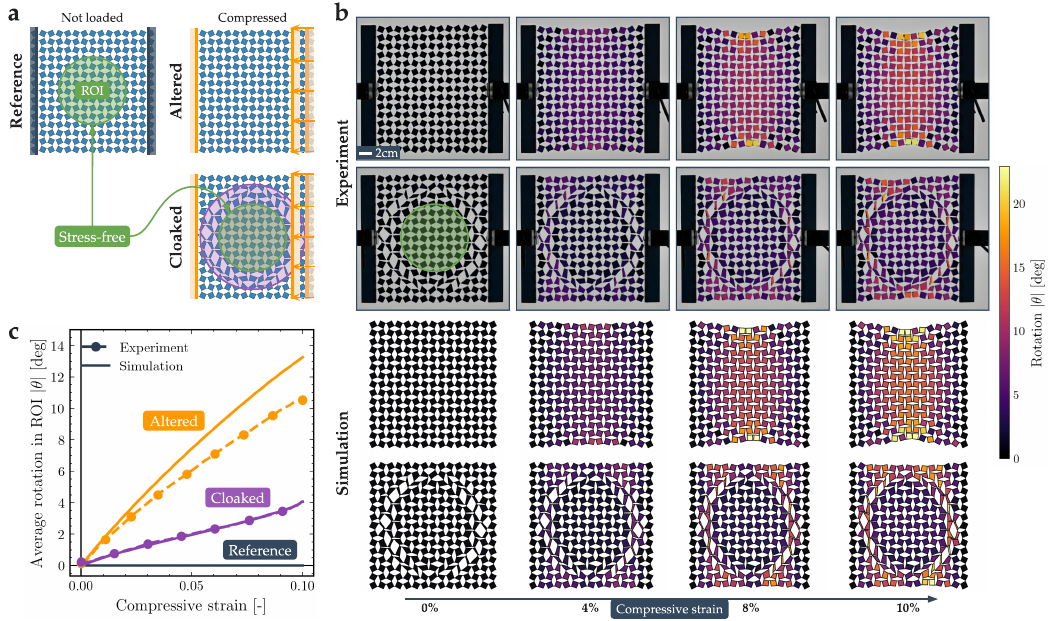}
    \caption{
        \textbf{Shielding quasi-static deformation.}
        (\textbf{a}) Cloaking design problem: (top-left) reference structure consisting of a $15\times15$-unit rotating-square metamaterial with spacing $s=\SI{15}{mm}$ and angle bias $\theta_0=\SI{20}{\degree}$, (top-right) the altered structure compressed at $\varepsilon=10\%$, and (bottom-right) the cloaked structure with a cloak size of $w_\text{cloak}=2s$ designed to create a stress-free circular region of radius $R_\text{ROI}=3s$ (green shaded area).
        (\textbf{b}) Experimental (top) and simulation (bottom) snapshots of the compressed reference and cloaked structures at $\varepsilon=0\%,\,4\%,\,8\%$, and $10\%$, from left to right.
        Colormap indicates the absolute value of the rotation of the units $\abs{\theta}$.
        (\textbf{c}) Evolution of the spatially-averaged absolute rotation $\abs{\theta}$ in the region of interest (green shaded area in panel \textbf{b}) as a function of compressive strain $\varepsilon$ for the cloaked (purple) and compressed reference (orange), and unloaded reference (black) structures.
        See also Movie~\ref{si-mov:movie_s2} for how such cloaks can protect fragile elements from mechanical stress.
    }
    \label{fig:fig_4}
\end{figure*}

While in Figures~\ref{fig:fig_2} and \ref{fig:fig_3} we designed a cloak to shield the boundary region of the metamaterial from a harmonic excitation applied at its center, we now shift focus to a quasi-static deformation applied at the boundaries and design a cloak that suppresses the resulting deformation in the central region of the metamaterial.
To demonstrate this concept, we again use the unloaded $15 \times 15$-unit rotating-square metamaterial as the reference configuration.
We perturb this configuration by applying a longitudinal strain of $\varepsilon = 10\%$, imposed through uniform horizontal displacements of all squares along the vertical edges, while constraining their vertical displacements and rotations (Figure~\ref{fig:fig_4}a, top-right).
The objective is to design a cloak that maintains the unstrained reference state in the center of the metamaterial, effectively shielding it from the applied macroscopic deformation.
To this end, we set the region of interest whose response we want to mimic to be the desired stress-free region (Figure~\ref{fig:fig_4}a, top-left), and use the mimicking function $g$ given by Equation~\eqref{eq:g_strain_energy} to minimize the difference between the strain energy of the reference (i.e. $\mW(\bu^\text{R})=0$) and cloaked structures.
Note that choosing the displacement-based mimicking function defined by Equation~\eqref{eq:g_displacement} would also be compatible with a stress-free target region, but it would also constrain the overall rigid-body motion, which is not a necessary requirement in this case.
Additionally, as the structure is compressed, any subpart of the domain experiences a non-zero average displacement irrespective of the metamaterial geometry, thus trying to minimize such displacement would lead to an ill-posed optimization problem.
The radius of the stress-free region is set to $R_\text{ROI}=3s$, and the cloak region consists of a small annular layer with width $w_\text{cloak}=2s$ (Figure~\ref{fig:fig_4}a, bottom-right).

The optimized cloak geometry identified by our algorithms is shown in Figure~\ref{fig:fig_4}b, alongside its physical realization.
It features a layer of high-aspect-ratio units and large voids.
In Figure~\ref{fig:fig_4}b we show experimental (top) and numerical (bottom) snapshots of both the cloak and reference structure under increasing compressive strain ($\varepsilon=0\%,\,4\%,\,8\%$, and $10\%$, see also Movie~\ref{si-mov:movie_s2}).
The quadrilateral units are color-coded by their absolute rotation $\abs{\theta}$, serving as a proxy for local stress.
In the reference structure, the applied compressive strain leads to large unit rotations near the center, signaling high stress.
In contrast, the cloaked structure exhibits only minor rotations in the central region, as the deformation is absorbed by the surrounding cloak and outer region.

To quantify performance, we track the evolution of the spatially-averaged absolute rotation $\abs{\theta}$ in the stress-free region (Figure~\ref{fig:fig_4}c).
For the reference structure $\abs{\theta}$ reaches $\SI{11}{\degree}$ at $\varepsilon = 10\%$, whereas for the cloaked structure it remains below $\SI{4}{\degree}$ up to $\varepsilon = 10\%$, corresponding to a 64\% reduction in average rotation.
Furthermore, the comparison between experimental and numerical data shows excellent agreement, confirming the effectiveness of our design strategy in discovering cloak designs with performance that reliably translates from simulation to experiment.

We further illustrate the effectiveness of the cloak effectiveness with a simple experimental demonstration: placing a piece of spaghetti in the protected region of the cloaked structure and another in the corresponding location of the reference.
When both structures are compressed under the same applied strain, the spaghetti in the cloaked structure remains intact, while the one in the reference breaks (see Movie~\ref{si-mov:movie_s2}).
This side-by-side comparison highlights the ability of the cloak to shield fragile elements from mechanical stress.

\subsection{Static cloaking of void inclusions}
\label{sec:cloaking_inclusion_static}
To further illustrate how our formulation extends cloaking into the nonlinear regime, we revisit the classical problem of concealing void inclusions.
We demonstrate that our method enables the design of nonlinear metamaterial cloaks capable of effectively hiding internal voids of arbitrary shape under static large-amplitude deformations.

\begin{figure*}
    \centering
    \includegraphics[width=\textwidth]{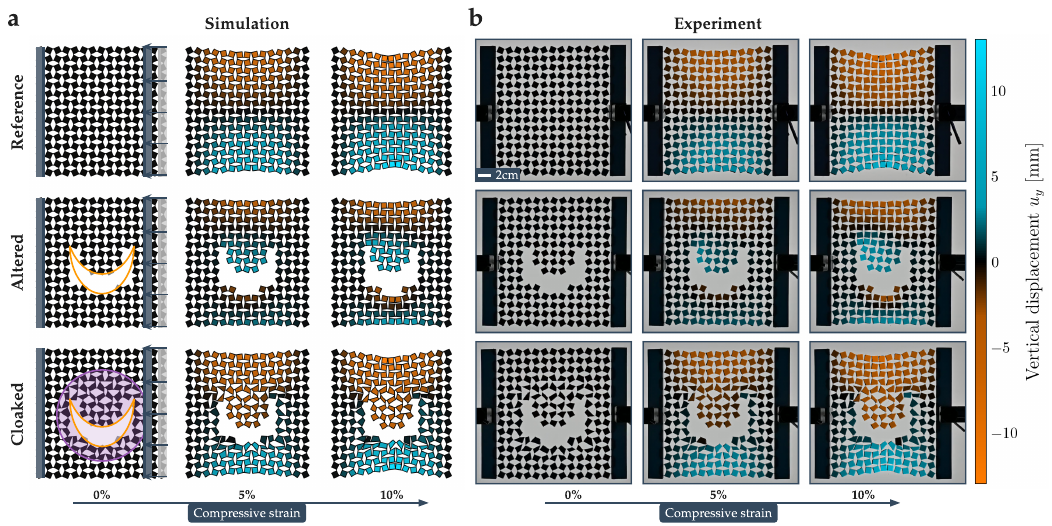}
    \caption{
        \textbf{Static cloaking of void inclusions.}
        (\textbf{a}) Simulation snapshots of the compressed reference ($s=\SI{15}{mm}$, $\theta_0=\SI{20}{\degree}$), altered (smile-shaped void inclusion), and optimized cloaked metamaterial structures at $\varepsilon=0\%,\,5\%$, and $10\%$ (left to right).
        In the optimized cloaked structure (bottom row), the smile-shaped void inclusion (orange shade) is surrounded by a circular cloak of radius $R_\text{cloak}=5s$ (purple shade).
        (\textbf{b}) Experimental snapshots of the three metamaterial structures at $\varepsilon=0\%,\,5\%$, and $10\%$ (left to right).
        For all snapshots, the colormap indicates the vertical displacement $u_y$.
        See also Movie~\ref{si-mov:movie_s3} for the DIC-tracked experimental videos of these structures.
    }
    \label{fig:fig_5}
\end{figure*}

As an example, we consider a $15 \times 15$-unit rotating-square metamaterial containing a smile-shaped void.
Comparing snapshots of the reference (intact) and altered (voided) metamaterials under a $\varepsilon=10\%$ compressive strain reveals that the intact structure exhibits significantly greater lateral contraction and also shows contact between units at the boundaries, unlike the voided one (Figure~\ref{fig:fig_5}a).
To counteract the influence of the void on the displacement field in the surrounding region, we apply the mimicking function defined in Equation~\eqref{eq:g_displacement} and solve the corresponding optimization problem of Equation~\eqref{eq:general_cloaking_problem}.
For a cloak region with radius $R_\text{cloak} = 5s$, our algorithm identify a cloak that successfully restores the original deformation field outside the cloak, despite experiencing substantial deformation and contact interactions within it (Figure~\ref{fig:fig_5}a).
The discovered metamaterial cloaked structure, along with the reference and voided structures, is fabricated and tested experimentally (Figure~\ref{fig:fig_5}b).
Excellent agreement is observed between experiments and simulations for all strain levels, confirming the effectiveness of our metamaterial model (see also Movie~\ref{si-mov:movie_s3}).
These results underscore the ability of our approach to design nonlinear cloaks that remain functional under large deformations and even contact interactions.
\begin{figure*}[htb]
    \centering
    \includegraphics[width=\textwidth]{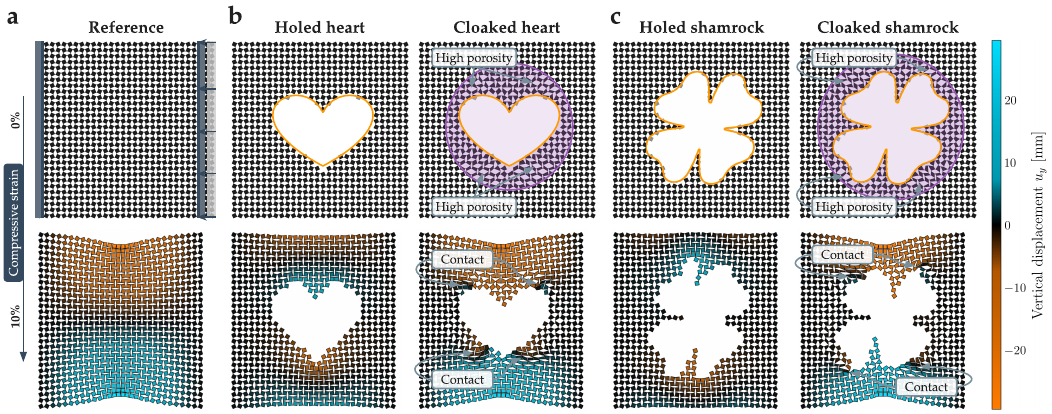}
    \caption{
        \textbf{Static cloaking of arbitrarily-shaped voids.}
        (\textbf{a}) Simulation snapshots of the intact $30\times30$-unit rotating-square metamaterial ($s=\SI{15}{mm}$, $\theta_0=\SI{20}{\degree}$) under compression at $\varepsilon=0\%$ (top) and $10\%$ (bottom).
        (\textbf{b}) Simulation snapshots for the heart-shaped void (left) and the corresponding optimized cloaked structure (right) at $\varepsilon=0\%$ (top) and $10\%$ (bottom).
        (\textbf{c}) Simulation snapshots for the shamrock-shaped void (left) and the corresponding optimized cloaked structure (right) at $\varepsilon=0\%$ (top) and $10\%$ (bottom).
        For all snapshots, the colormap indicates the vertical displacement $u_y$.
        See also Movie~\ref{si-mov:movie_s3} for simulation videos of these structures under compression.
    }
    \label{fig:fig_6}
\end{figure*}

Our cloaking framework supports voids of arbitrary shape.
Specifically, the geometry of the void can be defined either through direct parameterization or via a binary image input.
Figure~\ref{fig:fig_6} illustrates numerical results for two additional cloaked void shapes—a heart and a shamrock—embedded in a $30 \times 30$-unit metamaterial domain subjected to up to $10\%$ compressive strain.
The influence of void shape on the displacement field varies significantly: the heart-shaped void reduces transverse contraction (Figure~\ref{fig:fig_6}b) compared to the reference case (Figure~\ref{fig:fig_6}a), whereas the shamrock-shaped void induces transverse expansion (Figure~\ref{fig:fig_6}c).
In both cases, despite the substantial disruption caused by the inclusions, the optimized cloak effectively restores the original displacement field outside the cloaked region, recovering the characteristic negative Poisson's ratio contraction observed in the reference response.
Although each cloak is tailored to a specific void shape, Figures~\ref{fig:fig_6}b and \ref{fig:fig_6}c reveal a common structural motif: both cloaks incorporate regions near the void boundary that, in the undeformed state, exhibit high porosity, and when deformed, undergo large deformations with contact interactions leading to fully collapsed pores.
Remarkably, the cloaking effect is achieved using an annular region only $2s$ wide, which is significantly thinner than the void itself.
These results highlight the efficiency of our metamaterial design strategy in generating compact and effective cloaking solutions.

\subsection{Dynamic cloaking of void inclusions}
\label{sec:cloaking_inclusion_dynamic}
Thus far, our demonstrations of nonlinear cloaking have focused on  static or slowly varying loads.
However, our cloaking framework can be directly applied to scenarios involving to high-frequency excitations that generate the propagation of nonlinear waves.

\begin{figure*}[htb]
    \centering
    \includegraphics[width=\textwidth]{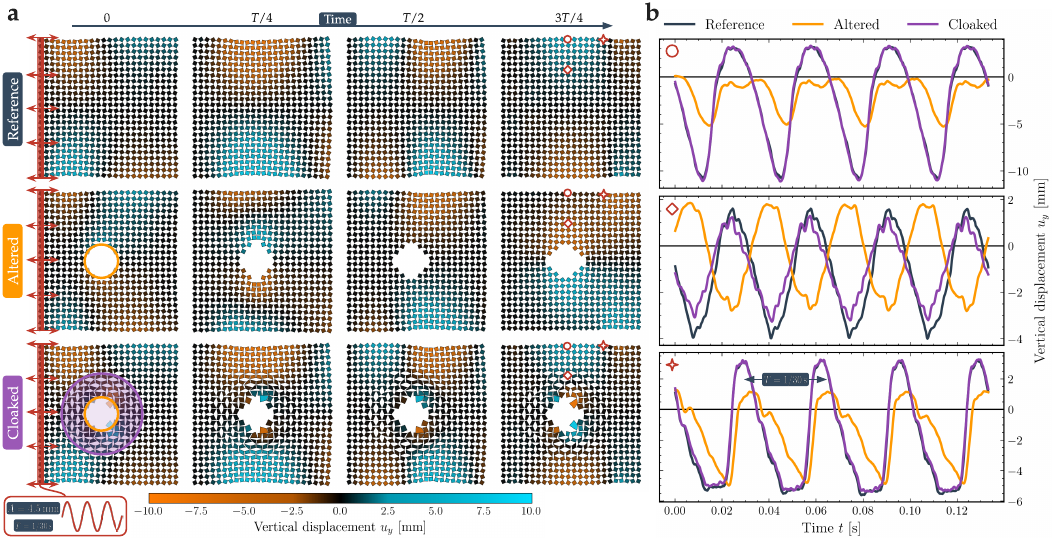}
    \caption{
        \textbf{Dynamic cloaking of void inclusions.}
        (\textbf{a}) Simulation snapshots of the intact $24\times24$-unit rotating-square metamaterial ($s=\SI{15}{mm}$, $\theta_0=\SI{20}{\degree}$) under harmonic excitation at the left boundary (top row), with a void inclusion of radius $R_\text{void}=3s$ (middle row), and with an optimized cloak of width $w_\text{cloak}=4s$ (bottom row).
        The applied harmonic excitation has frequency $f=\SI{30}{Hz}$ and amplitude $A=0.3s=\SI{4.5}{mm}$.
        The colormap indicates the vertical displacement $u_y$.
        (\textbf{b}) Time evolution of the vertical displacement field $u_y$ at three locations marked by circular, diamond, and four-pointed star red markers in panel \textbf{a}.
        The response of the intact structure (black) is compared with that of the altered structure with a void inclusion (orange) and the cloaked structure (purple).
        See also Movie~\ref{si-mov:movie_s4} for simulation videos of these structures under dynamic excitations.
    }
    \label{fig:fig_7}
\end{figure*}

To illustrate this, we consider a larger $24 \times 24$-unit rotating-square metamaterial domain that is harmonically excited at the left boundary along the $x$ direction with frequency $f = \SI{30}{Hz}$ and amplitude $A = 0.3s = \SI{4.5}{mm}$.
As shown by the numerical snapshots in the top row of Figure~\ref{fig:fig_7}a, such dynamic loading generates a large-amplitude nonlinear wave pattern propagating from left to right across the structure.
The nonlinear nature of the wave is confirmed by the presence of contact interactions between neighboring units on the bottom and top edges (see also Movie~\ref{si-mov:movie_s4}).
The introduction of a circular void inclusion of radius $R_\text{void}=3s$ significantly alters the wave patterns, in particular in terms of phase and amplitude of the traveling waves (Figure~\ref{fig:fig_7}a, middle row).
Furthermore, contact interactions at the bottom and top edges are completely suppressed due to the reduced lateral contraction caused by the void.
To neutralize the effect of the void on the wave pattern, we design a cloak with a width $w_\text{cloak}=4s$ by solving Equation~\eqref{eq:general_cloaking_problem}, using the displacement-based mimicking function $g$ of Equation~\eqref{eq:g_displacement}.
The bottom row of Figure~\ref{fig:fig_7}a reports snapshots of the response of the structure with the optimized cloak around the void.
By comparing the spatiotemporal evolution of the displacement field in the reference (intact), altered (voided), and cloaked structures, we find that the optimized cloak successfully suppresses the void wave scattering and restores the transmitted wave profile to closely match that of the intact structure, including the re-emergence of contact interactions at the top and bottom edges.
A quantitative comparison is provided in Figure~\ref{fig:fig_7}b, where we plot the time evolution of the vertical displacement field $u_y$ at three different locations marked by circular, diamond, and four-pointed star red markers in Figure~\ref{fig:fig_7}a.
The reported time series confirm that the cloaked structure (purple) closely follows the response of the intact structure (black), while the voided structure (orange) exhibits significant deviations in amplitude, phase, and shape of the waveform.
Notably, even at the cloak boundary (diamond marker), the response of the cloaked structure closely matches that of the intact structure, although some discrepancies arise due to the strong nonlinearity of the wave and the proximity to the cloak.

%

\section{Conclusions}
\label{sec:conclusions}
In summary, we have introduced a design strategy for mechanical cloaking in the nonlinear regime, grounded in a behavior-mimicking optimization framework. This generalized approach enables the creation of cloaks that neutralize the effects of both structural modifications and external disturbances.
By leveraging differentiable simulations, we systematically identified optimal cloaking architectures, which were subsequently fabricated and tested under both static and dynamic loading conditions.
Our results demonstrate that this strategy can be used to develop cloaks that shield regions from external excitations, conceal embedded inclusions, and create stress-free zones within metamaterial structures.
These capabilities open new directions for designing flexible structures whose responses are unaffected by internal heterogeneities, enabling mechanical decoupling of subdomains within a monolithic architecture.
Potential applications include vibration isolation, protection of sensitive components such as sensors, and the generation of controlled deformations in soft robotic systems with embedded components.

\section*{Data Availability Statement}
The experimental data, optimization data, and simulation results supporting the findings of this study are available on Zenodo at \href{https://doi.org/10.5281/zenodo.16952370}{doi.org/10.5281/zenodo.16952370}.
The source code is available on GitHub at \href{https://github.com/bertoldi-collab/MechanicalMetamaterialCloaks}{github.com/bertoldi-collab/MechanicalMetamaterialCloaks}.

\acknow{
    This work was supported by the Simons Collaboration on Extreme Wave Phenomena Based on Symmetries and by the National Science Foundation (NSF) under award no. 2118201, and by CNRS via IRP DynaMetaFlex.
    Thanks to Leon M. Kamp and Jackson K. Wilt for insightful discussions.
}
\showacknow{} 

\bibliography{nonlinear_cloaking} 

\end{document}

%% file: definitions.tex
\DeclarePairedDelimiter{\abs}{\lvert}{\rvert}
\DeclarePairedDelimiter{\norm}{\lVert}{\rVert}

\newcommand{\bu}{\bm u}

\newcommand{\bx}{\bm x}

\newcommand{\bU}{\bm U}

\newcommand{\bxi}{\bm \xi}

\newcommand{\mW}{\mathcal W}